\documentclass[11pt,letterpaper]{article}
\usepackage[letterpaper,textwidth=6.5in,textheight=8.5in,centering]{geometry}
\usepackage{titling}
\usepackage{bm}
\usepackage{natbib}
\usepackage{authblk}
\usepackage{xcolor}
\date{}
\onecolumn
\usepackage{amsmath, amssymb, amscd, amsthm, amsfonts}
\usepackage{graphicx}
\usepackage{hyperref}
\hypersetup{colorlinks=true,linkcolor=blue,citecolor=blue}
\usepackage{tikz}
\usetikzlibrary{arrows.meta}

\usepackage{subcaption}
\usepackage{placeins}

\usepackage{threeparttablex}
\usepackage{booktabs,multirow}
\usepackage{url} \usepackage{listings}
\makeatletter
\newcommand*{\indep}{\mathbin{\mathpalette{\@indep}{}}}
\newcommand*{\nindep}{\mathbin{\mathpalette{\@indep}{\not}}}
\newcommand*{\@indep}[2]{\sbox0{$#1\perp\m@th$}\sbox2{$#1=$}\sbox4{$#1\vcenter{}$}\rlap{\copy0}\dimen@=\dimexpr\ht2-\ht4-.2pt\relax
	\kern\dimen@
	{#2}\kern\dimen@
	\copy0 }

\makeatother
\theoremstyle{plain}
\newtheorem{theorem}{Theorem}
\newtheorem{lemma}{Lemma}
\newtheorem{proposition}{Proposition}
\newtheorem{corollary}{Corollary}
\theoremstyle{plain}
\newtheorem{assumption}{Assumption}
\theoremstyle{plain}

\theoremstyle{plain}

\theoremstyle{plain}

\newenvironment{sequation*}{\begin{equation*}\small}{\end{equation*}}

\newenvironment{tequation*}{\begin{equation*}\tiny}{\end{equation*}}

\usepackage{mathrsfs}

\usepackage{accents}

\graphicspath{{figure/}}
\usepackage{color}

\allowdisplaybreaks[4]

\renewcommand{\thesubfigure}{\Alph{subfigure}}

\usepackage{siunitx}

\allowdisplaybreaks[4]
\begin{document}

\title{Efficient causal survival inference under limited surrogate monitoring}
\author[1,\textdagger]{Zhenyue Du}
\author[2,\textdagger]{Weishang Kong}
\author[3]{Wei Li\thanks{Corresponding author. Email: \href{mailto:weilistat@ruc.edu.cn}{weilistat@ruc.edu.cn}}}
\affil[1]{Department of Mathematics and Statistics, University of Massachusetts Amherst, Amherst, MA 01003, U.S.A. \href{mailto:dzhenyue@umass.edu}{dzhenyue@umass.edu}}
\affil[2]{School of Statistics, Renmin University of China, 59 Zhongguancun Street, Beijing 100872, China\\ \href{mailto:Weishang_Kong@outlook.com}{Weishang\_Kong@outlook.com}}
\affil[3]{Center for Applied Statistics and School of Statistics, Renmin University of China, 59 Zhongguancun Street, Beijing 100872, China}
\begin{NoHyper}
\maketitle
\end{NoHyper}
\begingroup
\renewcommand{\thefootnote}{\fnsymbol{footnote}}
\makeatletter
\long\def\@makefntext#1{\makethanksmark #1}
\makeatother
\footnotetext[2]{Both authors contributed equally to this work.}
\endgroup

\begin{abstract}
Post-treatment surrogate endpoints can improve the
efficiency of causal survival effect estimation, but
their monitoring may end before follow-up for the
primary outcome. We investigate the impact of
limited recurrence monitoring on semiparametric
efficiency when mortality follow-up continues.
We derive the efficient influence function under
limited monitoring and establish an exact expression
for the efficiency gain from extending monitoring
between any two durations. 
The resulting efficiency comparison characterizes when
extending surrogate monitoring improves precision and
distinguishes these gains from those available with
a fully observed surrogate.
We develop a doubly robust estimator under limited
surrogate monitoring and establish asymptotic
normality and semiparametric efficiency for its
cross-fitted version under suitable conditions.
Simulations and an application to breast cancer
data illustrate the proposed method.
\end{abstract}


\vfill
\newpage

\section{Introduction}
\label{sec:introduction}

Time-to-event outcomes are primary endpoints in many clinical and
observational studies, with overall survival being a prominent example.
Studies often also record intermediate events, such as disease progression
or recurrence, that precede death and are informative about subsequent
survival. Together with death, these events define commonly used composite
endpoints, including progression-free survival and disease-free survival
\citep{buyse2007progression,sargent2005disease}. In practice, however,
intermediate events and the primary outcome need not be monitored for the
same duration. Recurrence assessments may end at a prespecified time while
mortality follow-up continues. This discrepancy affects how much
post-treatment event-history information is available for estimating a
causal survival effect. It also raises a practical statistical question:
how much additional precision can be gained by extending intermediate-event
monitoring, and how does this gain depend on the monitoring duration?

A substantial literature on surrogate endpoints examines whether an
intermediate outcome can substitute for the primary endpoint in evaluating
treatment effects. Beginning with the criterion of
\citet{prentice1989surrogate}, approaches to surrogate validation include
measures based on the proportion of treatment effect explained
\citep{freedman1992statistical,wang2002measure} and more flexible methods
\citep{parast2016robust,agniel2025robust}. Our objective is distinct.
We retain the causal effect on the primary survival outcome as the target
of inference and use intermediate events as auxiliary information, without
requiring them to be valid substitutes for the primary endpoint.

The use of auxiliary information to improve statistical
efficiency has a long history \citep{pepe1992inference}.
Several studies have investigated the use of auxiliary or
surrogate information to improve treatment effect estimation
when primary outcomes are partially missing
\citep{cheng2021robust,zeng2024continuous,
kallus2025role,hou2025efficient}.
For right-censored survival outcomes, intermediate-event
and longitudinal information has been incorporated through
various modelling and estimation approaches
\citep{finkelstein1994analysing,murray1996nonparametric,
flandre1996auxiliary,cook2001efficiency,
parast2014landmark,zee2015surrogate,
chen2023secondary,mukhin2026surrogates}.
These studies demonstrate the statistical value of auxiliary
information under different observation mechanisms.
However, they do not directly characterize how the
efficiency of causal survival effect estimation varies
when recurrence monitoring ends before mortality follow-up
and the monitoring cutoff changes. In this setting,
recurrence events occurring after the cutoff are not
recorded, even for individuals whose subsequent survival
is observed.

We consider composite time-to-event surrogates arising from
intermediate events and death, which exhibit a semicompeting-risks
structure. We allow monitoring of the intermediate event to
terminate at a prespecified time, while mortality follow-up may
continue. We derive the efficient influence function under limited
monitoring and characterize how the semiparametric efficiency
bound varies with the monitoring duration. In particular, we
establish an exact expression for the efficiency gain from extending
monitoring between any two durations. This expression relates the
gain to the additional prognostic information revealed by longer
monitoring and the primary-outcome censoring mechanism, thereby
characterizing when extending monitoring improves precision.
We also compare the resulting efficiency bounds with those
obtained without surrogate information, under shared censoring,
and with a fully observed surrogate.

Building on the efficient influence function under limited
monitoring, we develop a doubly robust estimator that combines
survival predictions based on the observed event history with
adjustment for treatment assignment and censoring.
To construct the required nuisance functions, we use an
illness--death representation together with a landmark
survival approach. We further develop a cross-fitted estimator
that accommodates flexible nuisance estimation and attains
the semiparametric efficiency bound under suitable conditions.
Simulation studies evaluate the finite-sample performance of
the proposed estimators and the efficiency gains across
monitoring durations. An application to the Molecular Taxonomy
of Breast Cancer International Consortium (METABRIC) study
illustrates how the duration of recurrence monitoring affects
the precision of estimated survival effects.

\section{Setup and notation} \label{sec:setup}

\subsection{Notation} \label{sub:notation}

Consider an observational or randomized study with $n$ independent individuals. Let $A\in\{0,1\}$ denote the binary treatment assignment, with $A=1$ for treatment and $A=0$ for control, and let $X$ denote a $p$-dimensional vector of pretreatment covariates. The primary endpoint is the time to a terminal event, denoted by $T$ (e.g., overall survival), which may be subject to right censoring by a censoring time $C$. Consequently, we observe $\tilde T=\min(T,C)$, together with the event indicator $\delta_T=I(T\le C)$. Following the potential outcomes framework \citep{rubin2005causal}, let $T(a)$ and $C(a)$ denote the potential primary event time and potential censoring time, respectively, under treatment level $a\in\{0,1\}$. Under consistency, $T=T(A)$, with an analogous relation holding for $C$.

Suppose that, in addition to the primary event time $T$, a post-treatment surrogate $S$ is available. The surrogate may be a clinical response measure, a biomarker, or a time-to-event endpoint such as progression-free survival or disease-free survival. Since the surrogate may itself be affected by treatment, let $S(a)$ denote its potential value under treatment level $a$, so that, under consistency, $S=S(A)$. We first consider the setting in which $S$ is
fully observed, with observed data
$O=(X,A,S,\tilde T,\delta_T)$. We then investigate the efficiency implications
of incomplete surrogate information, focusing on
limited monitoring of composite time-to-event
surrogates. Throughout, we assume that the relevant event-time and censoring distributions are absolutely continuous conditional on the relevant covariates, so that the corresponding conditional hazard functions are well defined.

Our primary estimand is the average treatment effect on survival at time $t$, $\theta(t) = E[ I\{T(1)\ge t\} - I\{T(0)\ge t\} ]$ for $ 0\le t\le t_{\max}$, where $t_{\max}$ denotes a prespecified finite analysis horizon. More generally, the proposed framework can be extended to causal estimands of the form $E\{\Gamma(T(1))-\Gamma(T(0))\}$, where $\Gamma(\cdot)$ is a known measurable transformation of the survival time. This class includes, for example, restricted mean survival time and other finite-horizon functionals of the survival distribution. The corresponding identification result and the  conditions on $\Gamma$ are provided in Section A.2 of the Supplementary Material. For each treatment level $a$, define $\theta_a(t) = E\left[I\{T(a)\ge t\}\right]$, so that $\theta(t)=\theta_1(t)-\theta_0(t)$. We first develop the identification and semiparametric efficiency theory for $\theta_a(t)$ at a generic treatment level $a$. The corresponding results for $\theta(t)$ are then obtained by taking the appropriate contrast of the treatment-specific results. Table~\ref{tab:notation} summarizes the key notation used throughout the paper.

\begin{table}[!htbp] \caption{Some key notation used throughout the paper.} \label{tab:notation} \centering \resizebox{0.98\textwidth}{!}{ \begin{tabular}{ll} \toprule Notation & Description\\ \midrule

$e_a(X)=\mathrm{pr}(A=a\mid X)$ & Propensity score. \\

$G(t\mid X,a)=\mathrm{pr}(C\ge t\mid X,A=a)$ & Survival function of the censoring time given $(X,A=a)$. \\

$G(t\mid S,X,a)=\mathrm{pr}(C\ge t\mid S,X,A=a)$ & Survival function of the censoring time given $(S,X,A=a)$. \\

$\mu(t\mid X,a)=\mathrm{pr}(T\ge t\mid X,A=a)$ & Survival function of the primary endpoint given $(X,A=a)$. \\

$\mu(t\mid S,X,a)=\mathrm{pr}(T\ge t\mid S,X,A=a)$ & Survival function of the primary endpoint given $(S,X,A=a)$. \\ $\pi(u\mid X,a)=\mathrm{pr}(S\ge u\mid X,A=a)$ & Survival function of the surrogate endpoint given $(X,A=a)$. \\

$\pi(t,u\mid X,a)=\mathrm{pr}(T\ge t,S\ge u\mid X,A=a)$ & Joint survival function of the primary endpoint and surrogate given $(X,A=a)$. \\

\bottomrule \end{tabular} } \end{table}

\subsection{Assumptions and identification} \label{subsec:assumptions}

To identify the parameter $\theta_a(t)$ from the observed data, we impose the following commonly-used assumptions in causal survival analysis.

\begin{assumption}[Unconfoundedness] \label{A1:unconfoundedness} $\{T(a),S(a),C(a)\}\perp A\mid X$. \end{assumption}

\begin{assumption}[Censoring at random] \label{A2:CAR} $T(a)\perp C(a)\mid S(a),X,A=a$. \end{assumption}

\begin{assumption}[Overlap] \label{A3:overlap} There exist $\epsilon_e>0$ and $\epsilon_C>0$ such that, $e_a(X)>\epsilon_e$, $G(t_{\max}\mid X,a)>\epsilon_C$, and $G(t_{\max}\mid S,X,a)>\epsilon_C$. \end{assumption}

Assumption~\ref{A1:unconfoundedness} states that treatment assignment is conditionally independent of the potential event time, surrogate, and censoring time given the observed covariates. In other words, the observed covariates are assumed to contain sufficient pre-treatment information to remove confounding due to non-random treatment assignment. Assumption~\ref{A2:CAR} requires that, conditional on the surrogate, covariates, and treatment assignment, the censoring time is independent of the primary event time. 
Assumption~\ref{A3:overlap} requires sufficient overlap in both the treatment assignment and censoring mechanisms. The first condition is the standard positivity assumption in causal inference \citep{rosenbaum1983central,imbens2015causal}, ensuring that every individual has a positive probability of receiving either treatment level. The second and third conditions require every individual to have a positive probability of remaining uncensored up to the maximum follow-up time, both before and after conditioning on the surrogate, thereby ruling out covariate or surrogate configurations under which the primary event is almost surely censored within the analysis window.

\begin{proposition} \label{T1:identification} Suppose Assumptions~\ref{A1:unconfoundedness}--\ref{A3:overlap} hold. Then \[ \theta_a(t) = E\bigg[ E\bigg\{ E\bigg( \frac{I(\tilde T\ge t)} {G(t\mid S,X,a)} \Bigm| S,X,A=a \bigg) \Bigm| X,A=a \bigg\} \bigg]. \] \end{proposition}

Proposition~\ref{T1:identification} identifies the counterfactual survival probability under treatment level $a$ by combining inverse probability of censoring weighting with surrogate information. The innermost conditional expectation corrects for right censoring and reconstructs the conditional survival probability given the surrogate, covariates, and treatment assignment. The second expectation averages over the conditional distribution of the surrogate given the covariates and treatment, thereby incorporating the surrogate as auxiliary information. Finally, the outer expectation averages over the covariate distribution to recover the marginal survival probability in the target population.

Proposition~\ref{T1:identification} assumes that the surrogate is fully observed. In practice, however, surrogate information may be incomplete. In particular, monitoring of intermediate events may terminate before follow-up for the primary survival outcome, leaving subsequent intermediate-event information unavailable even when mortality follow-up continues. This motivates our investigation of semiparametric efficiency under different observation schemes.  Figure~\ref{fig:three_tables} provides an overview of the three settings (I)--(III) considered throughout the paper.

\begin{figure}[htbp]
\renewcommand{\thesubfigure}{\Roman{subfigure}}
\centering

\newlength{\paneltableheight}
\setlength{\paneltableheight}{2.5cm}

\begin{subfigure}[t]{0.26\linewidth}
\centering
\resizebox{\linewidth}{\paneltableheight}{%
\begin{tabular}{ccccc}
\toprule
\(X\) & \(A\) & \(S\) & \(T\) & \(\delta_T\)\\
\midrule
\(\checkmark\) & \(\checkmark\) & -- & \(\checkmark\) & 1\\
\(\vdots\) & \(\vdots\) & \(\vdots\) & \(\vdots\) & \(\vdots\)\\
\(\checkmark\) & \(\checkmark\) & -- & \(\checkmark\) & 1\\
\(\checkmark\) & \(\checkmark\) & -- & ? & 0\\
\(\checkmark\) & \(\checkmark\) & -- & ? & 0\\
\(\vdots\) & \(\vdots\) & \(\vdots\) & \(\vdots\) & \(\vdots\)\\
\(\checkmark\) & \(\checkmark\) & -- & ? & 0\\
\bottomrule
\end{tabular}%
}
\caption{No surrogate}
\label{tab:setting1}
\end{subfigure}
\hfill
\begin{subfigure}[t]{0.28\linewidth}
\centering
\resizebox{\linewidth}{\paneltableheight}{%
\begin{tabular}{ccccc}
\toprule
\(X\) & \(A\) & \(S\) & \(T\) & \(\delta_T\)\\
\midrule
\(\checkmark\) & \(\checkmark\) & \(\checkmark\) & \(\checkmark\) & 1\\
\(\vdots\) & \(\vdots\) & \(\vdots\) & \(\vdots\) & \(\vdots\)\\
\(\checkmark\) & \(\checkmark\) & \(\checkmark\) & \(\checkmark\) & 1\\
\(\checkmark\) & \(\checkmark\) & \(\checkmark\) & ? & 0\\
\(\checkmark\) & \(\checkmark\) & \(\checkmark\) & ? & 0\\
\(\vdots\) & \(\vdots\) & \(\vdots\) & \(\vdots\) & \(\vdots\)\\
\(\checkmark\) & \(\checkmark\) & \(\checkmark\) & ? & 0\\
\bottomrule
\end{tabular}%
}
\caption{Fully observed}
\label{tab:setting2}
\end{subfigure}
\hfill
\begin{subfigure}[t]{0.28\linewidth}
\centering
\resizebox{\linewidth}{\paneltableheight}{%
\begin{tabular}{ccccc}
\toprule
\(X\) & \(A\) & \(S\) & \(T\) & \(\delta_T\)\\
\midrule
\(\checkmark\) & \(\checkmark\) & \(\checkmark\) & \(\checkmark\) & 1\\
\(\vdots\) & \(\vdots\) & \(\vdots\) & \(\vdots\) & \(\vdots\)\\
\(\checkmark\) & \(\checkmark\) & \(\checkmark\) & ? & 0\\
\(\checkmark\) & \(\checkmark\) & ? & \(\checkmark\) & 1\\
\(\vdots\) & \(\vdots\) & \(\vdots\) & \(\vdots\) & \(\vdots\)\\
\(\checkmark\) & \(\checkmark\) & ? & ? & 0\\
\bottomrule
\end{tabular}%
}
\caption{Limited monitoring}
\label{tab:setting3}
\end{subfigure}

\caption{
Illustration of three observation schemes.
A ``$\checkmark$'' indicates an observed value, ``--''
information not collected, and 
``?'' an unknown exact value.
}
\label{fig:three_tables}
\end{figure}

\section{Semiparametric efficiency analysis} \label{sec:efficiency}

\subsection{Fully observed surrogates} \label{subsec:setting2}

We first consider the setting introduced in Section~\ref{sec:setup}, where the surrogate is fully observed for every individual, while the primary survival outcome may be subject to right censoring. We derive the efficient influence function for the treatment-specific survival probability $\theta_a(t)$ under the nonparametric model. The derivation uses the counting-process representation for right-censored survival data \citep{tsiatis2006semiparametric}. Let $N_C(u)=I(\tilde T\le u,\delta_T=0)$ denote the censoring counting process and $Y(u)=I(\tilde T\ge u)$ the corresponding at-risk process. Define the censoring martingale $M_C(u;S,X,a) = N_C(u) - \int_0^u Y(v)\, d\Lambda_C(v\mid S,X,A=a)$, where $\Lambda_C(\cdot\mid S,X,A=a)$ denotes the conditional cumulative hazard function of the censoring time. The following theorem characterizes the efficient influence function for $\theta_a(t)$.

\begin{theorem} \label{thm:EIF2} Under Assumptions~\ref{A1:unconfoundedness}--\ref{A3:overlap}, the efficient influence function for $\theta_a(t)$ under Setting (II) is 
\begin{align*}
\psi_{II}(t;a,O)
&=
\frac{I(\tilde T\ge t)I(A=a)}
     {e_a(X)G(t\mid S,X,a)}+\mu(t\mid X,a)-\theta_a(t)
-
\frac{I(A=a)}{e_a(X)}
\bigg\{
\mu(t\mid X,a)
\\
&\quad
-
\int_0^t
\frac{\mu(t\mid S,X,a)}
     {\mu(u\mid S,X,a)G(u\mid S,X,a)}
\,dM_C(u;S,X,a)
\bigg\}.
\end{align*}
\end{theorem}

Theorem~\ref{thm:EIF2} characterizes the efficient influence function when the surrogate is fully observed. The surrogate enters through the conditional survival and censoring functions in the inverse probability weighting and martingale augmentation terms.
The resulting efficient influence function provides the basis for assessing the efficiency gains from incorporating surrogate information. We next consider limited monitoring of intermediate events, where recurrence monitoring may terminate before follow-up for the primary  outcome.

\subsection{Limited recurrence monitoring} \label{subsec:setting3}

We now consider a time-to-event surrogate subject to limited monitoring. A common example is a composite endpoint, such as progression-free or disease-free survival, defined as the time to disease progression or recurrence, or death, whichever occurs first. Let $R$ denote the time to progression or recurrence and $T$ the time to death. The composite surrogate time is defined as $S=\min(R,T)$, with the convention $R=\infty$ when death occurs without a preceding progression or recurrence event. We assume that the two events do not occur simultaneously. In practice, monitoring for progression or recurrence may terminate before follow-up for the primary survival outcome. Let $b\ge 0$ denote the prespecified end of surrogate monitoring, and define $C_b=\min(C,b)$ as the effective censoring time for the surrogate. The observed surrogate time and its event indicator are $\tilde S_b=\min(S,C_b)$ and $\delta_{S,b}=I(S\le C_b)$, respectively. Thus, the observed data under Setting~(III) are $O_b=(X,A,\tilde S_b,\delta_{S,b}, \tilde T,\delta_T)$.

Unlike the usual shared-censoring setting, limited monitoring allows the surrogate time to remain incompletely observed even when the primary event time is known. In particular, when death occurs after surrogate monitoring has ended, the exact surrogate time may remain unknown. This observation pattern is illustrated in Setting~(III) of Fig.~\ref{fig:three_tables}. When $b=0$, no surrogate information is collected. For a fixed target time $t$, if $b\ge t$, the available surrogate history up to $t$ coincides with that under shared censoring, although observations beyond $t$ may differ.

Because the complete surrogate time is not necessarily observed, the identification formula in Proposition~\ref{T1:identification} and the efficient influence function derived in Section~\ref{subsec:setting2} cannot be applied directly. To characterize the efficiency bound under limited monitoring, we impose the following additional assumption.

\begin{assumption}[Surrogate--censoring independence] \label{A4:CARS} $S(a)\perp C(a)\mid X,A=a$. \end{assumption}

Together with Assumption~\ref{A2:CAR}, Assumption~\ref{A4:CARS} implies $C(a)\perp\{S(a),T(a)\}\mid X,A=a$. Thus, conditional on baseline covariates and treatment assignment, the censoring time is independent of the surrogate and primary event times. In particular, $G(u\mid S,X,a)=G(u\mid X,a)$, so that the censoring martingale defined in Section~\ref{subsec:setting2} simplifies to $M_C(u;X,a) = N_C(u) - \int_0^u Y(v)\, d\Lambda_C(v\mid X,A=a)$. 
The monitoring cutoff $b$ determines the available surrogate information, but does not alter the censoring mechanism for the primary survival outcome.

For $0\le u\le t$, let $\mathcal H_u^b$ denote the surrogate event history up to $u\wedge b$, generated by $\{I(S\le v):0\le v\le u\wedge b\}$ on the risk set $\{T\ge u\}$. Define the conditional survival prediction $Q_b(t,u;S,X,a) = \mathrm{pr}(T\ge t\mid T\ge u,\mathcal H_u^b, X,A=a)$. Here, the dependence on $S$ is understood through its observed history up to $u\wedge b$, rather than its complete value. Using the notation introduced in Section~\ref{sec:setup}, this prediction admits the following explicit representation: 
\begin{equation*}
Q_b(t,u;S,X,a)=
\begin{cases}
\displaystyle
\frac{\pi(t,u\mid X,a)}
{\pi(u\mid X,a)},
& u\le b,\ S\ge u,\\[8pt]
\displaystyle
\frac{\mu(t\mid S,X,a)}
{\mu(u\mid S,X,a)},
& S\le b,\ S<u,\\[8pt]
\displaystyle
\frac{\pi(t,b\mid X,a)}
{\pi(u,b\mid X,a)},
& b<u\le t,\ S>b.
\end{cases}
\end{equation*}
The ratios are defined whenever their denominators
are positive. On the risk set $\{T\ge u\}$,
the three cases correspond to no surrogate event
by $u$ during monitoring, an intermediate event
observed before $u$, and no surrogate event
by the monitoring cutoff $b$, respectively.
In the last case, subsequent intermediate-event
information is unavailable, although follow-up
for the primary outcome continues.

\begin{theorem} \label{thm:EIF3} Under Assumptions~\ref{A1:unconfoundedness}--\ref{A4:CARS}, the efficient influence function for $\theta_a(t)$ under Setting~(III) with monitoring duration $b$ is 
\begin{equation*}
\begin{aligned}
\psi(t;a,O_b)
={}&\mu(t\mid X,a)-\theta_a(t)+
\frac{I(A=a)}{e_a(X)}
\left\{
\frac{I(\tilde T\ge t)}{G(t\mid X,a)}
-\mu(t\mid X,a)
\right.
\\
&\quad\left.
+
\int_0^t
\frac{Q_b(t,u;S,X,a)}
{G(u\mid X,a)}
\,dM_C(u;X,a)
\right\}.
\end{aligned}
\end{equation*}
\end{theorem}

Theorem~\ref{thm:EIF3} characterizes the efficient
influence function under limited recurrence monitoring.
The monitoring duration enters through the
history-dependent survival prediction
$Q_b(t,u;S,X,a)$, which uses only the surrogate
history available up to $u\wedge b$.
The censoring martingale remains defined in terms
of the primary-outcome censoring time $C$, since
mortality follow-up may continue after surrogate
monitoring ends.
Thus, the efficient influence function accounts
for intermediate events that may remain unobserved
after the monitoring cutoff.
The following corollary shows that
Theorem~\ref{thm:EIF3} recovers the efficient
influence functions under two familiar observation
schemes.

\begin{corollary} \label{cor:EIF-boundary} Under the conditions of Theorem~\ref{thm:EIF3}, the following results hold for a fixed target time $t$.

\noindent \textup{(a)} When $b=0$, \[ Q_b(t,u;S,X,a) = \frac{\mu(t\mid X,a)}{\mu(u\mid X,a)}, \qquad 0\le u\le t. \] Consequently, $\psi(t;a,O_b)$ reduces to the efficient influence function under Setting~(I).

\noindent \textup{(b)} When $b\ge t$, on the risk set $\{T\ge u\}$, \[ \begin{aligned} Q_b(t,u;S,X,a) =I(S\ge u) \frac{\pi(t,u\mid X,a)}{\pi(u\mid X,a)} + I(S<u) \frac{\mu(t\mid S,X,a)} {\mu(u\mid S,X,a)}, \qquad 0\le u\le t. \end{aligned} \] Consequently, $\psi(t;a,O_b)$ reduces to the efficient influence function under the usual shared-censoring scheme. \end{corollary}

Corollary~\ref{cor:EIF-boundary} connects the proposed framework to two familiar observation schemes. When $b=0$, no surrogate information is available, and the efficient influence function reduces to that based on the censored primary outcome alone, as studied by \citet{bai2013doubly}. When $b\ge t$, it reduces to the shared-censoring setting, which was also investigated independently in contemporaneous work by \citet{mukhin2026surrogates}. For intermediate monitoring durations, the proposed framework incorporates the partially observed surrogate history through $Q_b(t,u;S,X,a)$.

\subsection{Efficiency comparison} \label{subsec:efficiency_gain}

We now compare the semiparametric efficiency bounds for the average treatment effect $\theta(t)=\theta_1(t)-\theta_0(t)$. Since $\theta(t)$ is a linear contrast of the treatment-specific survival probabilities, its efficient influence function is the difference between the corresponding treatment-specific efficient influence functions.

Under Setting~(III), the efficiency bound depends on the surrogate monitoring duration $b$. Let $\psi(t;O_b) = \psi(t;1,O_b)-\psi(t;0,O_b)$ and $ \mathcal V_b(t) = \operatorname{var}\{\psi(t;O_b)\}$. Let $\mathcal V_I(t)$ and $\mathcal V_{II}(t)$ denote the efficiency bounds under Settings~(I) and (II), respectively. By Corollary~\ref{cor:EIF-boundary}, $\mathcal V_0(t)=\mathcal V_I(t)$ and $\mathcal V_b(t)=\mathcal V_t(t)$ for $b\ge t$. We write $\mathcal V_{\mathrm{sc}}(t)=\mathcal V_t(t)$ for the efficiency bound under shared censoring. The following theorem characterizes the efficiency gain from extending surrogate monitoring.

\begin{theorem} \label{T3:efficiency} Under Assumptions~\ref{A1:unconfoundedness}--\ref{A4:CARS}, for any $0\le b_1\le b_2\le t$, \begin{equation}\label{eq:efficiency-monitoring} \begin{aligned} \resizebox{0.93\linewidth}{!}{$\displaystyle \mathcal V_{b_1}(t)-\mathcal V_{b_2}(t) = \sum_{a=0}^1 \int_0^t E\bigg[ \frac{\lambda_C(u\mid X,a)\mu(u\mid X,a)} {e_a(X)G(u\mid X,a)} E\left\{ (Q_{b_2}-Q_{b_1})^2 \mid T\ge u,X,A=a \right\} \bigg]du,$} \end{aligned} \end{equation} where $Q_{b_j}=Q_{b_j}(t,u;S,X,a)$ for $j=1,2$. Consequently, $\mathcal V_I(t) \ge \mathcal V_b(t) \ge \mathcal V_{\mathrm{sc}}(t) \ge \mathcal V_{II}(t)$ for $0\le b\le t$, where the comparison with Setting~(II) is made under the same underlying composite-surrogate model. \end{theorem}

Theorem~\ref{T3:efficiency} shows that extending surrogate monitoring cannot increase the semiparametric efficiency bound. The efficiency gain is determined by the additional prognostic information revealed through longer monitoring, as measured by the squared difference between the corresponding conditional survival predictions. The gain is strictly positive when the additional surrogate history changes these predictions on risk sets with positive censoring intensity. In particular, taking $b_1=0$ and $b_2=t$ in \eqref{eq:efficiency-monitoring} yields the efficiency gain from surrogate information under shared censoring relative to Setting~(I). The additional efficiency gain from fully observing the surrogate, relative to shared censoring, is 
\begin{equation*}
\begin{aligned}
\resizebox{0.99\linewidth}{!}{$\displaystyle  \mathcal V_{\mathrm{sc}}(t)
-\mathcal V_{II}(t)
=
\sum_{a=0}^1
\int_0^t
E\bigg[
\frac{\lambda_C(u\mid X,a)\pi(u\mid X,a)}
{e_a(X)G(u\mid X,a)}
\operatorname{var}
\{
\mu(t\mid S,X,a)
\mid S\ge u,X,A=a\}
\bigg]du.$}
\end{aligned}
\end{equation*}
This expression shows that the additional gain depends on the prognostic heterogeneity in the complete surrogate time among individuals who remain surrogate-event-free at time $u$. Together, these results distinguish the efficiency gains from extending surrogate monitoring and from observing the surrogate completely.

\section{Doubly robust estimation} \label{sec:estimation}

\subsection{Estimator construction} \label{sub:construction}

We now construct an estimator for $\theta(t)$ based on the efficient influence function under Setting~(III), allowing surrogate monitoring to terminate before follow-up for the primary survival outcome. An estimator under Setting~(II) can be constructed analogously using its corresponding efficient influence function.

The efficient influence function in Theorem~\ref{thm:EIF3} involves several nuisance functions associated with the joint distribution of $(S,T)$. To facilitate estimation, we adopt an illness--death representation of the underlying event process. For $u\le b$, let $\lambda_{01}(u\mid X,a)$ and $\lambda_{02}(u\mid X,a)$ denote the transition hazards from the initial state to the intermediate event and directly to death, respectively. For individuals whose intermediate event occurs at time $s\le b$, let $\lambda_{12}(u\mid s,X,a)$ denote the subsequent death hazard for $u>s$. This hazard may depend on the time of the intermediate event, allowing a non-Markov specification. Since intermediate events are no longer monitored after $b$, $\lambda_{01}$ and $\lambda_{02}$ are generally not separately identifiable beyond $b$ without additional restrictions.

For individuals who remain surrogate-event-free
at the monitoring cutoff, define
$\nu(t\mid X,a)
=\mathrm{pr}(T\ge t\mid S>b,X,A=a)$
for $t\ge b$, with $\nu(b\mid X,a)=1$.
We collect the nuisance functions as
$\eta=\{\lambda_{01},\lambda_{02},\lambda_{12},
\nu,e_a,\lambda_C\}$,
where $\lambda_C$ denotes the censoring hazard.
The transition hazards are defined over their
respective observable event histories.
For $(j,k)\in\{(0,1),(0,2)\}$ and $u\le b$,
define
$\Lambda_{jk}(u\mid X,a)
=\int_0^u\lambda_{jk}(v\mid X,a)\,dv$.
For $s\le b$ and $t>s$, let
$\Lambda_{12}(t\mid s,X,a)
=\int_s^t\lambda_{12}(v\mid s,X,a)\,dv$,
and define
$\Lambda_C(u\mid X,a)
=\int_0^u\lambda_C(v\mid X,a)\,dv$.
The following lemma expresses the survival
quantities entering the efficient influence
function in Theorem \ref{thm:EIF3} in terms of the nuisance functions in $\eta$.

\begin{lemma} \label{lem:nuisance_decomposition} Let $0\le u\le t\le t_{\max}$ and $r=t\wedge b$. Then the following results hold.

\begin{enumerate}

\item[(a)] For $0\le u\le b$, the surrogate survival function is $\pi(u\mid X,a) = \exp\{ -\Lambda_{01}(u\mid X,a) -\Lambda_{02}(u\mid X,a) \}$. \item[(b)] For $0\le u\le r$, the joint survival function satisfies $\pi(t,u\mid X,a) = \pi(r\mid X,a) \nu(t\vee b\mid X,a) + \int_u^r \pi(s\mid X,a) \lambda_{01}(s\mid X,a) \exp\{-\Lambda_{12}(t\mid s,X,a)\} \,ds$. \item[(c)] For  $s\le b$, if $t\leq s$, then $\mu(t\mid S=s,X,a)=1$; if $t>s$, then \[ \mu(t\mid S=s,X,a)=\frac{\lambda_{01}(s\mid X,a)} {\lambda_{01}(s\mid X,a)+ \lambda_{02}(s\mid X,a)} \exp\{-\Lambda_{12}(t\mid s,X,a)\}. \]

\item[(d)] The  survival function of $T$ is $\mu(t\mid X,a) = \pi(t,0\mid X,a)$. Moreover,  \[ \begin{aligned} Q_b(t,u;S,X,a) = \begin{cases} \displaystyle \frac{\pi(t,u\mid X,a)} {\pi(u\mid X,a)}, & u\le b,\ S\ge u,\\[9pt] \displaystyle \exp\left\{ -\int_u^t \lambda_{12}(v\mid S,X,a)\,dv \right\}, &  S\le b,\ S<u,\\[9pt] \displaystyle \frac{\nu(t\mid X,a)} {\nu(u\mid X,a)}, & b<u\le t,\ S>b. \end{cases} \end{aligned} \]

\item[(e)] The censoring survival function satisfies $G(u\mid X,a) = \exp\{-\Lambda_C(u\mid X,a)\}$. \end{enumerate} \end{lemma}

Lemma~\ref{lem:nuisance_decomposition}
relates the required survival quantities to
the transition hazards and the landmark
survival function $\nu$.
In particular, when $t>b$, the joint survival
function in part (b) involves $\nu(t\mid X,a)$.
Although
$\nu(t\mid X,a)
=\pi(t,b\mid X,a)/\pi(b\mid X,a)$,
the joint survival probability
$\pi(t,b\mid X,a)$ cannot be determined from
the transition hazards before $b$ alone.
Since recurrence is no longer monitored after
$b$, the subsequent transition hazards out
of the recurrence-free state are not separately
identifiable.
Instead, $\nu$ can be estimated directly
from mortality follow-up among individuals
known to satisfy $S>b$.


We next construct an estimator based on the efficient influence function in Theorem~\ref{thm:EIF3}. To emphasize its dependence on the nuisance functions, let $\varphi_a(t;O_b;\eta)$ denote the uncentred estimating function obtained from $\psi(t;a,O_b)$ by removing the centring term $-\theta_a(t)$ and evaluating the nuisance functions at $\eta$. At the true nuisance functions $\eta_0$, $E\{\varphi_a(t;O_b;\eta_0)\}=\theta_a(t)$. Let $\mathbb P_n g=n^{-1}\sum_{i=1}^n g(O_{b,i})$ denote the empirical mean operator, and let $\hat\eta$ denote estimators of the nuisance functions. We define $\hat\theta_a(t) = \mathbb P_n\{\varphi_a(t;O_b;\hat\eta)\}$, and estimate the treatment effect by $\hat\theta(t)=\hat\theta_1(t)-\hat\theta_0(t)$.

The estimator accommodates generic nuisance estimators $\hat\eta$. In practice, the nuisance functions may be estimated using parametric or semiparametric working models. For example, the propensity score may be estimated by logistic regression, while the transition and censoring hazards may be estimated using Cox proportional hazards models. The transition hazards $\lambda_{01}$ and $\lambda_{02}$ are estimated using events observed before $b$, whereas $\lambda_{12}$ can be estimated using subsequent mortality follow-up among individuals whose intermediate event was observed during monitoring. The latter hazard may depend on the time of entry into State~1, allowing a non-Markov specification. The conditional survival function $\nu$ may be estimated using a landmark survival model among individuals known to be surrogate-event-free at $b$. The corresponding cumulative baseline hazards can be estimated using the Breslow estimator. By Lemma~\ref{lem:nuisance_decomposition}, estimates of the nuisance functions in $\eta$ determine all survival quantities required to evaluate $\hat\theta(t)$.

\subsection{Double robustness} \label{sub:DR}

The proposed estimator enjoys a double robustness property with respect to two groups of nuisance functions: the event-time models $\{\lambda_{01},\lambda_{02},\lambda_{12},\nu\}$ and the treatment-assignment and censoring mechanisms $\{e_a,\lambda_C\}$.

\begin{theorem} \label{T4:DR} Under Assumptions~\ref{A1:unconfoundedness}--\ref{A4:CARS} and the regularity conditions in Section~A.9 of the Supplementary Material, the estimator $\hat\theta_a(t)$ is consistent for $\theta_a(t)$ if either of the following conditions holds:

\begin{enumerate} \item[(a)] The event-time nuisance functions $\lambda_{01}$, $\lambda_{02}$, $\lambda_{12}$, and $\nu$ are consistently estimated over their respective relevant domains;

\item[(b)] The propensity score $e_a$ and the censoring hazard $\lambda_C$ are consistently estimated. \end{enumerate} \end{theorem}

The double robustness property can be understood through the bias of the uncentred estimating function. Let $\eta^*$ denote the probability limit of the nuisance estimators under possible misspecification, and define $\mathcal B_a(t;\eta^*) = E\{\varphi_a(t;O_b;\eta^*)\} -\theta_a(t)$. We use a superscript $*$ to denote the survival and propensity functions induced by $\eta^*$, and write $\Delta^*(t,u\mid X,a) = E[ I(T\ge u) \{Q_b^*(t,u;S,X,a) -Q_b(t,u;S,X,a)\} \mid X,A=a]$. As shown in the proof of Theorem~\ref{T4:DR}, the bias admits the decomposition \begin{equation*} \begin{aligned} \mathcal B_a(t;\eta^*) ={}& E\bigg[ \frac{e_a(X)}{e_a^*(X)} \int_0^t \frac{G(u\mid X,a)} {G^*(u\mid X,a)} \Delta^*(t,u\mid X,a) \{d\Lambda_C(u\mid X,a) -d\Lambda_C^*(u\mid X,a)\} \bigg] \\ &+ E\bigg[ \frac{e_a^*(X)-e_a(X)} {e_a^*(X)} \{\mu^*(t\mid X,a) -\mu(t\mid X,a)\} \bigg]. \end{aligned} \end{equation*}

This decomposition reveals the product-error
structure underlying double robustness.
The first term involves errors in the
history-dependent survival prediction and
the censoring mechanism, while the second
involves errors in the marginal survival
function and the propensity score.
Under condition~(a) of Theorem~\ref{T4:DR},
Lemma~\ref{lem:nuisance_decomposition} implies
that $\mu^*=\mu$ and $Q_b^*=Q_b$, so both
bias terms vanish.
Under condition~(b), both terms also vanish,
regardless of misspecification of the
event-time models.

\subsection{Cross-fitted estimation} \label{sub:crossfit}

To accommodate flexible, data-adaptive estimation of the nuisance functions, we employ cross-fitting. Fix an integer $K\ge 2$ and partition $\{1,\ldots,n\}$ into $K$ approximately equal-sized folds $\mathcal I_1,\ldots,\mathcal I_K$. Let $n_k=|\mathcal I_k|$ and $\mathcal I_{-k}=\{1,\ldots,n\}\setminus \mathcal I_k$. For each fold $k$, let $\hat\eta^{-k} = \{ \hat\lambda_{01}^{-k}, \hat\lambda_{02}^{-k}, \hat\lambda_{12}^{-k}, \hat\nu^{-k}, \hat e_a^{-k}, \hat\lambda_C^{-k} \}$ denote the nuisance estimators trained using observations in $\mathcal I_{-k}$. The corresponding estimators of the marginal
survival function $\mu$, the conditional survival
prediction $Q_b$, and the censoring survival
function $G$ are obtained from
Lemma~\ref{lem:nuisance_decomposition}.

Write $\hat\varphi_a^{-k}(t;O_b) =\varphi_a(t;O_b;\hat\eta^{-k})$ for the uncentred estimating function evaluated at these nuisance estimates. The cross-fitted estimator is defined by $\hat\theta_a^{\mathrm{cf}}(t) = K^{-1}\sum_{k=1}^K\, \mathbb P_{n_k}\{\hat\varphi_a^{-k}(t;O_b)\}$, where $\mathbb P_{n_k}g=n_k^{-1}\sum_{i\in\mathcal I_k}g(O_{b,i})$. The corresponding treatment-effect estimator is $\hat\theta^{\mathrm{cf}}(t) = \hat\theta_1^{\mathrm{cf}}(t) - \hat\theta_0^{\mathrm{cf}}(t)$. By evaluating each observation using nuisance estimators trained without its own fold, cross-fitting permits flexible nuisance estimation without imposing Donsker-type complexity conditions, subject to suitable convergence conditions \citep{chernozhukov2018double}.

We next establish asymptotic normality for $\hat\theta^{\mathrm{cf}}(t)$. For a measurable function $g(O_b)$, write $\|g\|_2=\{E[g^2(O_b)]\}^{1/2}$. Expectations involving estimated nuisance functions are understood conditionally on the corresponding training sample. For  $k=1,\ldots,K$, define $\Delta_a^{-k}(t,u\mid X) = E[ I(T\ge u) \{\hat Q_b^{-k}(t,u;S,X,a) -Q_b(t,u;S,X,a)\} \mid X,A=a ]$. Also define the signed censoring-error measure \[ d\mathcal D_{C,a}^{-k}(u\mid X) = \frac{G(u\mid X,a)} {\hat G^{-k}(u\mid X,a)} \left\{ d\Lambda_C(u\mid X,a) -d\hat\Lambda_C^{-k}(u\mid X,a) \right\}. \] Consider the following remainder terms: \begin{align} r_{1,a}^{-k} &= E\Big[ \big| \{\hat e_a^{-k}(X)-e_a(X)\} \{\hat\mu^{-k}(t\mid X,a) -\mu(t\mid X,a)\} \big| \Big], \label{eq:r1} \\ r_{2,a}^{-k} &= E\bigg[ \Big| \int_0^t \Delta_a^{-k}(t,u\mid X) \,d\mathcal D_{C,a}^{-k}(u\mid X) \Big| \bigg]. \label{eq:r2} \end{align}

\begin{theorem} \label{T_AN} Suppose Assumptions~\ref{A1:unconfoundedness}--\ref{A4:CARS} and the regularity conditions in Section~A.10 of the Supplementary Material hold. For each $a\in\{0,1\}$ and $k=1,\ldots,K$, assume the following conditions:

\begin{enumerate} \item[(a)] There exist constants $c_e,c_G>0$ such that, with probability tending to one, $\hat e_a^{-k}(X)\ge c_e$, and $\inf_{0\le u\le t} \hat G^{-k}(u\mid X,a)\ge c_G$ on the relevant support. All estimated conditional survival ratios  satisfy the required boundedness conditions.

\item[(b)] The estimating functions satisfy $\| \varphi_a(t;\cdot;\hat\eta^{-k}) - \varphi_a(t;\cdot;\eta_0) \|_2=o_p(1)$.

\item[(c)] The remainder terms satisfy $r_{j,a}^{-k}=o_p(n^{-1/2})$ for $j=1,2$. \end{enumerate} Then \[  n^{1/2}\{ \hat\theta^{\mathrm{cf}}(t)-\theta(t) \} \overset{d}{\longrightarrow} N\{0,\mathcal V_b(t)\}, \] where $\mathcal V_b(t) = E[ \{ \psi(t;1,O_b)-\psi(t;0,O_b) \}^2]$. Consequently, $\hat\theta^{\mathrm{cf}}(t)$ attains the semiparametric efficiency bound under Setting~(III) with monitoring duration $b$. \end{theorem}

The conditions in \eqref{eq:r1}--\eqref{eq:r2}
reflect the product-error structure of the bias
decomposition in Section~\ref{sub:DR}.
Specifically, $r_{1,a}^{-k}$ controls the
interaction between errors in the propensity
score and marginal survival function, whereas
$r_{2,a}^{-k}$ controls the interaction between
errors in the history-dependent survival
prediction $Q_b$ and the censoring mechanism.
By Lemma~\ref{lem:nuisance_decomposition},
the errors in $\mu$ and $Q_b$ can be related
to estimation errors in the event-time nuisance
functions $\lambda_{01}$, $\lambda_{02}$,
$\lambda_{12}$, and $\nu$.
Under suitable regularity conditions, the
remainder terms can therefore be controlled
by products of the primitive nuisance estimation errors in $\eta$.
Explicit sufficient product-rate conditions
are provided in
Lemma~S5
of the Supplementary Material.
These conditions permit nuisance estimators
to converge at slower than the parametric
$n^{-1/2}$ rate, provided that the required
product-rate conditions hold.

For variance estimation, define $\hat\psi^{-k}(t;O_b) = \hat\varphi_1^{-k}(t;O_b) - \hat\varphi_0^{-k}(t;O_b) - \hat\theta^{\mathrm{cf}}(t)$. A consistent estimator of the asymptotic variance is $\hat{\mathcal V}_b(t) = K^{-1}\sum_{k=1}^K \mathbb P_{n_k} \{ (\hat\psi^{-k}(t;O_b))^2 \}$. An asymptotic $(1-\alpha)$ confidence interval for $\theta(t)$ is $ \hat\theta^{\mathrm{cf}}(t) \pm z_{1-\alpha/2} \{\hat{\mathcal V}_b(t)/n\}^{1/2}$, where $z_{1-\alpha/2}$ is the $(1-\alpha/2)$ quantile of the standard normal distribution.

For subsequent comparisons, we also consider the cross-fitted estimator under Setting~(I), constructed from the efficient influence function for right-censored survival data without surrogate information. We denote this estimator by $\hat\theta^{\mathrm{cf,I}}(t)$, which serves as the no-surrogate benchmark for evaluating efficiency gains under limited recurrence monitoring.

\section{Simulation} \label{sec:simulation}

\subsection{Data-generating mechanism}

We conduct simulation studies to evaluate the finite-sample performance of the proposed estimators. We first examine the double robustness property in Theorem~\ref{T4:DR} under limited surrogate monitoring using parametric working models. We then evaluate the cross-fitted estimators with flexible nuisance estimation and compare their efficiency across monitoring durations. Finally, we evaluate the efficiency bounds using the true nuisance functions to examine the efficiency comparisons established in Theorem~\ref{T3:efficiency}.

All experiments share the same data-generating mechanism. The baseline covariates $X_1$ and $X_2$ are generated independently from $\operatorname{Unif}(-1,1)$. Conditional on $X=(X_1,X_2)^\top$, treatment $A\in\{0,1\}$ follows a Bernoulli distribution with \[ \operatorname{logit}\{\mathrm{pr}(A=1\mid X)\} = 0.9X_1-0.7X_2+1.1X_1X_2 +0.9(X_1^2-1/3). \] The event process follows the illness--death structure introduced in Section~\ref{sub:construction}, with transitions $0\to1$, $0\to2$, and $1\to2$, representing recurrence, death without prior recurrence, and death following recurrence, respectively. The transition hazards are specified as exponential regression models with treatment-specific covariate effects. In particular, the $1\to2$ hazard also depends on the time of entry into the intermediate state, allowing for a non-Markov illness--death process. Censoring times are generated from an exponential regression model, independently of the recurrence and death times conditional on treatment and baseline covariates. The complete transition and censoring hazard specifications are provided in Section~B of the Supplementary Material.

The target estimand is the average treatment effect on survival at $t=5$, $ \theta(5) $. We consider sample size $n\in\{1000,2000,4000\}$. All Monte Carlo experiments use $2000$ replications per sample size. The true parameter value and the efficiency bounds are evaluated by numerical integration under the data-generating mechanism, as detailed in Section~B of the Supplementary Material.

\subsection{Double robustness} \label{subsec:sim_parametric}

We first investigate the double robustness property under Setting~(III), with $b=3<t=5$.  The propensity score is estimated by logistic regression, while the transition and censoring hazards are estimated using exponential regression models. The $1\to2$ transition model additionally includes recurrence entry time as a covariate. For this parametric experiment, the landmark survival function $\nu$ is derived from the fitted transition models by extending their parametric specifications beyond the monitoring cutoff. This construction is specific to the parametric simulation and is not required by the proposed general estimation framework.

We consider four nuisance-model specification scenarios: (i) all nuisance models are correctly specified; (ii) the event nuisance group is misspecified, while the propensity score and censoring models are correctly specified; (iii) the propensity score and censoring models are misspecified, while the event nuisance group is correctly specified; and (iv) both nuisance groups are misspecified. Correctly specified models include the covariate terms used in the data-generating mechanism. Misspecified hazard models include only an intercept and treatment, while the misspecified propensity model includes only an intercept. The misspecified $1\to2$ model also omits recurrence entry time.

We also include an oracle estimator evaluated using the true nuisance functions. For the fitted estimators, standard errors are calculated using a stacked sandwich estimator based on the treatment-effect estimating equation and the score equations for the propensity score, transition, and censoring models, thereby accounting for nuisance parameter estimation. Table~\ref{tab:sim} summarizes the results. The proposed estimator exhibits small empirical bias in scenarios (i)--(iii), consistent with the double robustness property established in Theorem~\ref{T4:DR}. The estimated standard errors generally agree with the Monte Carlo standard deviations, and coverage probabilities are close to the nominal $95\%$ level, with some finite-sample deviations. Both the empirical standard deviations and estimated standard errors decrease as the sample size increases. In contrast, when both nuisance groups are misspecified in scenario (iv), the empirical bias remains approximately $-0.018$ across the three sample sizes. As the sampling variability decreases, this persistent bias leads to increasingly pronounced undercoverage, with coverage probabilities decreasing from $92.6\%$ at $n=1000$ to $87.5\%$ at $n=4000$.

\begin{table}[!htbp]
\centering
\caption{Simulation results under different nuisance-model specifications with $b=3$ and $t=5$. Bias, MCSD, SE, and CP denote empirical bias, Monte Carlo standard deviation, mean estimated standard error, and empirical coverage probability of the nominal $95\%$ confidence interval, respectively. PS denotes propensity score. The oracle estimator uses the true nuisance functions.}
\label{tab:sim}
\begin{tabular}{ c l S[table-format=-1.4] S[table-format=1.4] S[table-format=1.4] S[table-format=2.1] } \hline {$n$} & Scenario & {Bias} & {MCSD} & {SE} & {CP (\%)} \\ \hline 1000 & Oracle                    & -0.0002 & 0.0445 & 0.0442 & 95.2 \\ & (i) All correct           & -0.0003 & 0.0452 & 0.0439 & 95.1 \\ & (ii) Event wrong          & -0.0005 & 0.0521 & 0.0492 & 93.9 \\ & (iii) PS/censor wrong     & -0.0008 & 0.0439 & 0.0437 & 94.9 \\ & (iv) All wrong            & -0.0191 & 0.0454 & 0.0443 & 92.6 \\ \addlinespace 2000 & Oracle                    &  0.0005 & 0.0317 & 0.0312 & 95.4 \\ & (i) All correct           &  0.0004 & 0.0317 & 0.0312 & 95.1 \\ & (ii) Event wrong          &  0.0001 & 0.0358 & 0.0349 & 95.0 \\ & (iii) PS/censor wrong     & -0.0001 & 0.0310 & 0.0309 & 95.4 \\ & (iv) All wrong            & -0.0185 & 0.0319 & 0.0313 & 90.6 \\ \addlinespace 4000 & Oracle                    &  0.0006 & 0.0224 & 0.0221 & 94.9 \\ & (i) All correct           &  0.0006 & 0.0224 & 0.0221 & 94.6 \\ & (ii) Event wrong          &  0.0006 & 0.0247 & 0.0248 & 95.5 \\ & (iii) PS/censor wrong     &  0.0005 & 0.0223 & 0.0218 & 94.6 \\ & (iv) All wrong            & -0.0180 & 0.0224 & 0.0222 & 87.5 \\ \hline \end{tabular}
\end{table}

\subsection{Flexible nuisance estimation} \label{subsec:sim_ml}

We next evaluate the proposed estimators using flexible nuisance estimation. For limited monitoring, the landmark survival function $\nu$ is estimated directly using a piecewise exponential model among individuals who remain uncensored and in State~0 at time $b$. Individuals with an observed recurrence before $b$ contribute to estimation of the $1\to2$ transition model. Further implementation details are provided in Section~B of the Supplementary Material.

To compare the proposed estimators across different monitoring schemes, we construct the corresponding observed datasets from the same underlying realization of $(X,A,S,T,C)$ in each Monte Carlo replication. Under Setting~(III), surrogate information is observed up to $C\wedge b$, while follow-up for the primary endpoint continues until $T\wedge C$. We consider monitoring cutoffs $b\in\{0,2,3.5,5\}$, together with the fully observed surrogate under Setting~(II). The choice $b=0$ corresponds to Setting~(I), and $b=5$ corresponds to shared censoring through the target time.

Table~\ref{tab:sim_grf} summarizes the results. The proposed estimators exhibit small empirical bias across the monitoring schemes, while both the Monte Carlo standard deviations and estimated standard errors decrease with sample size. The estimated standard errors generally agree with the Monte Carlo standard deviations, with closer agreement at larger sample sizes. At $n=4000$, the absolute bias does not exceed $0.0006$, and coverage probabilities range from $95.0\%$ to $95.6\%$. We assess the empirical variance reduction associated with surrogate monitoring by comparing the squared Monte Carlo standard deviations across monitoring schemes, using $b=0$ as the reference. The empirical variance decreases as monitoring duration increases at all three sample sizes. At $n=4000$, monitoring to $b=2$, $3.5$, and $5$ reduces the empirical variance by approximately $12.7\%$, $14.1\%$, and $15.7\%$, respectively, relative to no surrogate information. The additional gain from extending monitoring beyond $b=2$ is relatively modest. The fully observed surrogate yields the largest variance reduction, approximately $39.3\%$. These findings are consistent with the efficiency improvements established in Theorem~\ref{T3:efficiency}.

\begin{table}[!htbp]
\centering
\caption{Simulation results for the cross-fitted estimators under different surrogate monitoring schemes. Bias, MCSD, SE, and CP denote empirical bias, Monte Carlo standard deviation, mean estimated standard error, and $95\%$ confidence interval coverage, respectively. VR denotes the percentage reduction in Monte Carlo variance relative to $b=0$. Full denotes the fully observed surrogate under Setting~(II).}
\label{tab:sim_grf}
\begin{tabular}{ c c S[table-format=-1.4] S[table-format=1.4] S[table-format=1.4] S[table-format=2.1] S[table-format=2.1] } \hline {$n$} & Monitoring & {Bias} & {MCSD} & {SE} & {CP (\%)} & {VR (\%)} \\ \hline 1000 & $b=0$     & -0.0006 & 0.0623 & 0.0599 & 95.5 &  0.0 \\ & $b=2$     &  0.0016 & 0.0565 & 0.0541 & 95.3 & 17.8 \\ & $~~b=3.5$ &  0.0014 & 0.0554 & 0.0535 & 95.7 & 21.0 \\ & $b=5$     &  0.0001 & 0.0549 & 0.0514 & 95.7 & 22.3 \\ & Full      & -0.0002 & 0.0473 & 0.0442 & 95.9 & 42.3 \\ \addlinespace 2000 & $b=0$     & -0.0004 & 0.0374 & 0.0378 & 95.3 &  0.0 \\ & $b=2$     &  0.0004 & 0.0340 & 0.0345 & 95.6 & 17.3 \\ & $~~b=3.5$ &  0.0004 & 0.0337 & 0.0341 & 95.7 & 18.6 \\ & $b=5$     &  0.0003 & 0.0334 & 0.0332 & 95.4 & 20.0 \\ & Full      &  0.0003 & 0.0283 & 0.0284 & 95.5 & 42.7 \\ \addlinespace 4000 & $b=0$     &  0.0004 & 0.0249 & 0.0255 & 95.6 &  0.0 \\ & $b=2$     &  0.0004 & 0.0233 & 0.0233 & 95.3 & 12.7 \\ & $~~b=3.5$ &  0.0006 & 0.0231 & 0.0231 & 95.5 & 14.1 \\ & $b=5$     &  0.0006 & 0.0229 & 0.0227 & 95.0 & 15.7 \\ & Full      &  0.0005 & 0.0194 & 0.0194 & 95.6 & 39.3 \\ \hline \end{tabular}
\end{table}

\begin{figure}[!htbp] \centering \includegraphics[width=0.72\textwidth] {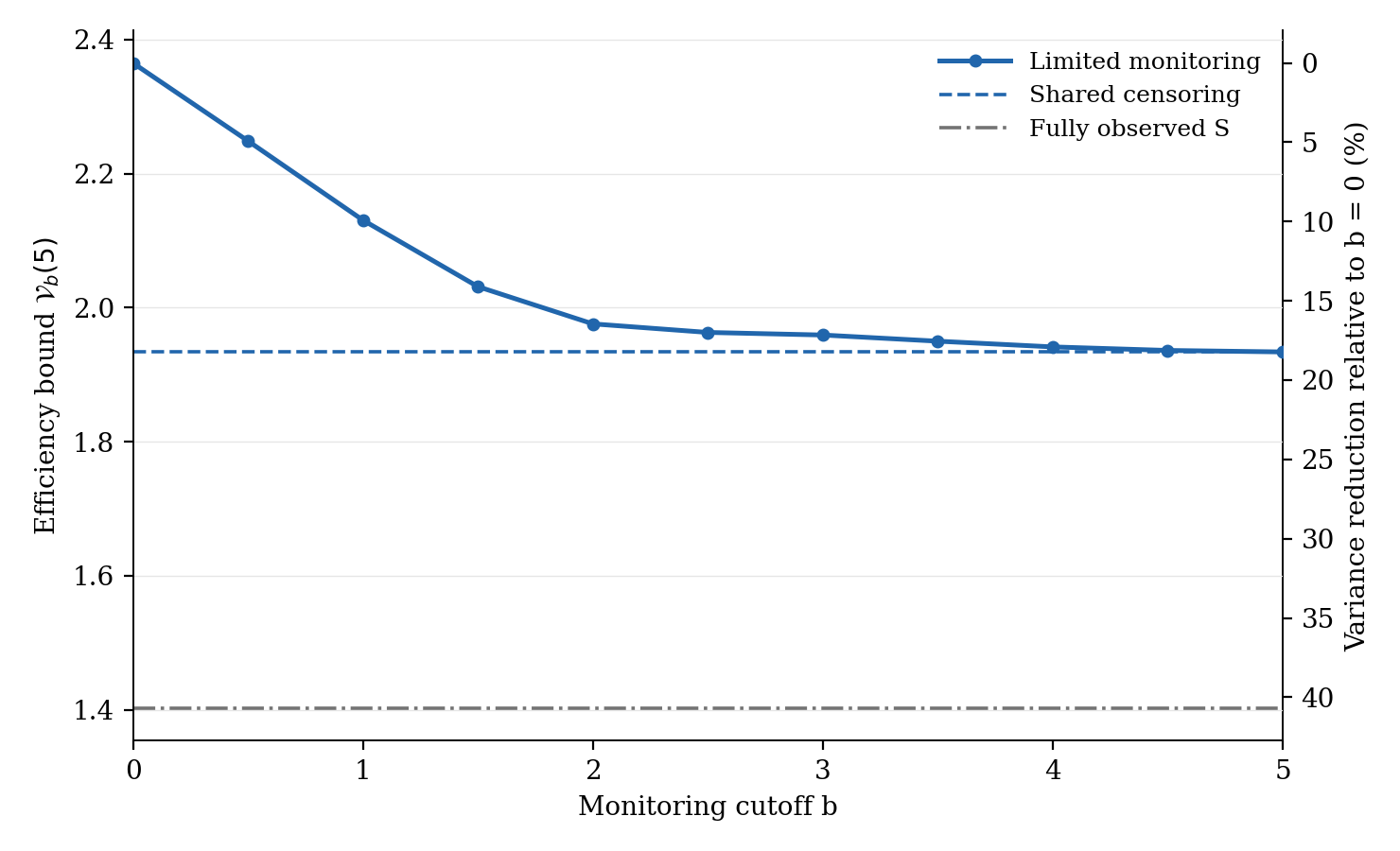} \caption{ Semiparametric efficiency bound $\mathcal V_b(5)$ evaluated using the true nuisance functions. The right axis indicates the percentage reduction in the efficiency bound relative to no surrogate information ($b=0$). } \label{fig:monitoring_bound}

\end{figure}

We further examine the effect of monitoring duration by evaluating the semiparametric efficiency bound $\mathcal V_b(5)$ using the true nuisance functions. We vary $b$ from $0$ to $5$ in increments of $0.5$ and calculate the corresponding efficiency bounds by numerical integration. The efficiency bound under the fully observed surrogate is evaluated separately. Figure~\ref{fig:monitoring_bound} shows that the efficiency bound decreases as monitoring duration increases, with most of the reduction occurring before $b=2$. Relative to no surrogate information, the theoretical variance reductions are approximately $16.5\%$, $17.5\%$, and $18.2\%$ for $b=2$, $3.5$, and $5$, respectively. The curve becomes relatively flat after $b=2$, indicating diminishing additional efficiency gains from extending monitoring under this data-generating mechanism. The fully observed surrogate yields a substantially larger variance reduction of approximately $40.7\%$. This additional gain arises because, under shared censoring, surrogate information beyond the individual censoring time remains unavailable, even when monitoring continues through the target time.

\section{Application to the METABRIC Dataset} \label{sec:realdata}

We illustrate the proposed method using data from the METABRIC study \citep{curtis2012genomic,pereira2016somatic}, which includes molecular profiling and clinical information for 2509 breast cancer patients. Previous work has used METABRIC to investigate the causal effects of hormone therapy and radiotherapy on recurrence and overall survival \citep{yu2025exploring}. We focus on chemotherapy and examine its effect on long-term survival, using disease recurrence as auxiliary information.

Specifically, we use relapse-free survival (RFS) and overall survival  records. The primary endpoint $T$ is time to death from any cause, and the composite surrogate is $S=\min(R,T)$, where $R$ denotes time to recurrence. Because the original RFS records include disease-specific death, we incorporate deaths from any cause to construct the composite endpoint. Both endpoints are assumed to share a common right-censoring time $C$. We define $A=1$ for patients who received chemotherapy and $A=0$ otherwise, adjusting for age at diagnosis, tumour size, number of positive lymph nodes, tumour grade, oestrogen receptor status, and type of breast surgery. After preprocessing, the analysis includes $1796$ patients, of whom $378$ received chemotherapy and $1418$ did not.

Our target estimand is $\theta(t)$, the average effect of chemotherapy on the survival probability at time $t$. We focus on survival probabilities between $60$ and $180$ months (5--15 years) after diagnosis. This interval is clinically relevant for evaluating long-term breast cancer survival and persistent recurrence risk \citep{colleoni2016annual,pan2017twenty,rueda2019dynamics}. We apply the proposed cross-fitted estimator, incorporating recurrence information through the illness--death structure. The nuisance functions are estimated using penalized spline models. For each target time $t$, we compare surrogate monitoring cutoffs $b=0$, $0.1t$, $0.2t$, $0.4t$, and $t$, while preserving the original follow-up for the primary survival outcome. The case $b=0$ corresponds to Setting~(I), which uses no surrogate information, whereas $b=t$ corresponds to shared censoring through the target time. These monitoring schemes are constructed by artificially restricting surrogate information in the same observed cohort.

\begin{figure}[htbp] 
\centering \includegraphics[width=\textwidth] {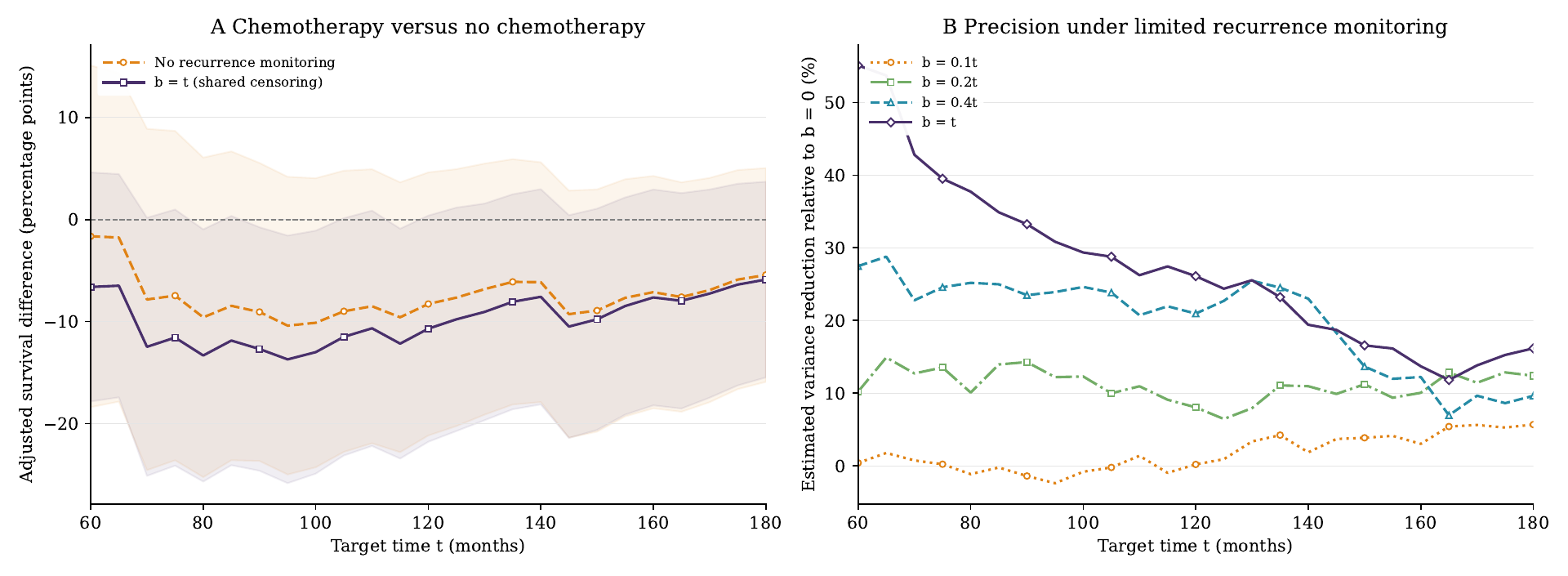} 
\caption{ 
(A) Estimated effect of chemotherapy on overall survival probability for $b=0$ and $b=t$ over $t\in[60,180]$ months. Shaded regions show pointwise $95\%$ confidence intervals. (B) Relative variance reduction for $b=0.1t$, $0.2t$, $0.4t$, and $t$ compared with estimation without recurrence information, expressed as a percentage. } 
\label{fig:metabric}


\end{figure}

Figure~\ref{fig:metabric}(A) shows that the estimated effects under $b=0$ and $b=t$ follow similar patterns and remain negative throughout the analysis period, with most pointwise confidence intervals including zero. Incorporating recurrence information yields narrower confidence intervals. Figure~\ref{fig:metabric}(B) shows that the estimated precision gains depend on monitoring duration. The gains under $b=0.1t$ are small, while monitoring through $b=0.4t$ generally yields larger gains than monitoring through $b=0.2t$ over the earlier and middle portions of the analysis period. The gains under $b=0.4t$ and $b=t$ are similar over part of the analysis period, although monitoring through $b=t$ provides larger gains at earlier target times. These results suggest that limited recurrence monitoring can capture a meaningful proportion of the available precision gain, qualitatively consistent with Fig.~\ref{fig:monitoring_bound}.

Table~\ref{tab:metabric_t120} provides a more detailed comparison at $t=120$ months (10 years after diagnosis), a clinically relevant time point for assessing long-term breast cancer survival. The estimated treatment effects are broadly similar and negative under all monitoring schemes, with all pointwise $95\%$ confidence intervals including zero. Compared with using no surrogate information, monitoring through $b=0.2t$, $0.4t$, and $t$ reduces the estimated variance by $8.0\%$, $21.0\%$, and $26.1\%$, respectively. Thus, monitoring through $b=0.4t$ captures a substantial proportion of the estimated precision gain obtained with monitoring through the target time.

\begin{table}[htbp]
\centering
\caption{
Estimated effects of chemotherapy at $t=120$ months.
Relative variance is the estimated variance under each monitoring scheme divided by that under $b=0$.
}
\label{tab:metabric_t120}
\begin{tabular}{ccccc}
\hline
Monitoring & Treatment effect & SE & $95\%$ CI
& Relative variance \\
\hline
$b=0~~~~$    & -0.0827 & 0.0658 & $[-0.2116,0.0462]$ & 1.000 \\
$b=0.1t$ & -0.0913 & 0.0657 & $[-0.2201,0.0374]$ & 0.998 \\
$b=0.2t$ & -0.1098 & 0.0631 & $[-0.2334,0.0137]$ & 0.920 \\
$b=0.4t$ & -0.1123 & 0.0585 & $[-0.2269,0.0023]$ & 0.790 \\
$b=t~~~~$    & -0.1069 & 0.0565 & $[-0.2177,0.0039]$ & 0.739 \\
\hline
\end{tabular}
\end{table}

\section{Discussion} \label{sec:discussion}

In this paper, we develop a framework for improving the efficiency of treatment effect estimation with censored survival outcomes by leveraging post-treatment surrogate information under limited monitoring. The framework characterizes the efficiency gains from extending surrogate monitoring while follow-up for the primary endpoint continues. The current formulation considers a single intermediate event and a fixed monitoring cutoff, with surrogate monitoring restricted by both the cutoff and the random censoring time governing follow-up for the primary endpoint. Extending the methodology to general multistate or recurrent-event processes would allow more complex post-treatment event histories to be exploited for efficiency improvement. Another possible extension is to allow distinct random censoring times for the surrogate and primary endpoints. Accommodating patient-specific monitoring schedules that depend on observed event histories would further broaden the applicability of the proposed framework.



\section*{Supplementary material} 

The supplementary material contains proofs of all theoretical results and additional simulation details.

\bibliographystyle{abbrvnat}
\bibliography{ref}

\end{document}